\documentclass[aps,prb, twocolumn, superscriptaddress]{revtex4-1}
\usepackage{graphicx}
\usepackage{amsmath}
\usepackage{wasysym}
\usepackage{mathrsfs}
\usepackage{times}

\usepackage{enumerate}
\newcommand{\beq}{\begin{equation}}
\newcommand{\eeq}{\end{equation}}
\newcommand{\bea}{\begin{eqnarray}}
\newcommand{\eea}{\end{eqnarray}}

\newcommand\NLSM{NL$\sigma$M}

\def\ket#1{{\left|#1\right\rangle}}

\def\br{\mathbf{r}}
\def\dipole{\boldsymbol{\mathfrak{p}}}

\begin{document}
\draft
\title{Microscopic Theory of a Quantum Hall Ising Nematic: Domain Walls and Disorder}
\author{Akshay Kumar}
\affiliation{Department of Physics, Princeton
University, Princeton, NJ 08544, USA}
\author{S.  A. Parameswaran}
\email{sidp@berkeley.edu}
\affiliation{Department of Physics, University of California, Berkeley, CA 94701, USA}
\author{S. L. Sondhi}
\affiliation{Department of Physics, Princeton
University, Princeton, NJ 08544, USA}

\date{\today}
\begin{abstract}
We study the the interplay between spontaneously broken valley symmetry and spatial disorder in multivalley semiconductors in the quantum Hall regime. In cases where valleys have  anisotropic electron dispersion a previous long-wavelength analysis [Phys. Rev. B {\bf 82}, 035428 (2010)] identified two new phases exhibiting the QHE. The first is the quantum Hall Ising nematic (QHIN), a phase with long-range orientational order manifested in macroscopic transport anisotropies. The second is the quantum Hall random-field paramagnet (QHRFPM) that emerges when the Ising ordering is disrupted by quenched disorder, characterized by a domain structure with a distinctive response to a valley symmetry-breaking strain field. Here we provide a more detailed microscopic analysis of the QHIN, which allows us to (i) estimate its Ising ordering temperature; (ii)  study its domain-wall excitations, which play a central role in determining its properties; and (iii) analyze its response to quenched disorder from impurity scattering, which gives an estimate for domain size in the descendant QHRFPM. Our results are directly applicable to AlAs heterostructures, although their qualitative aspects inform other ferromagnetic QH systems,   such as Si(111) heterostructures and bilayer graphene with trigonal warping.
\end{abstract}
\maketitle


\section{Introduction}
Two-dimensional electron gases (2DEGs) placed in high magnetic fields exhibit a multitude of phases associated with the quantum Hall effect (QHE).\cite{Pinzuk95} Particularly interesting in this context are situations in which quantum Hall ordering is accompanied by the breaking of internal symmetries -- such as the global symmetries associated with the electron spin,\cite{PhysRevB.47.16419} or valley\cite{PhysRevLett.55.433,Rasolt:1986p1} or layer\cite{Moon:1995p1} pseudospin. The resulting broken-symmetry state, termed a quantum Hall ferromagnet, possesses in addition to the topological order common to all quantum Hall states a distinctive set of phenomena relating to the low-energy pseudospin degrees of freedom. These include charged skyrmions, finite-temperature phase transitions, and Josephson-like effects, to name a few.

Recent experimental\cite{Shkolnikov:2005p1,Eng:2007p1,Padmanabhan:2010p1,Gokmen:2010p1,PhysRevB.84.125319,KaneValleys2012} and theoretical work has focused on the case in which the symmetry in question is between the different valleys (i.e., conduction band minima) of a semiconductor. In previous work~\cite{Abanin:2010p1} involving two of the present authors, it was noted that a generic feature of such multivalley systems is that the point-group symmetries act simultaneously on the internal valley pseudospin index and on the spatial degrees of freedom. This linking of pseudospin and space has significant consequences at ``ferromagnetic" filling factors, such as $\nu=1$:
\begin{enumerate}[{\it (i)}]
\item in the absence of disorder, pseudospin ferromagnetism onsets via an Ising-type finite-temperature transition and is necessarily accompanied by broken rotational symmetry, corresponding to nematic order. The resulting state at $T=0$, dubbed the quantum Hall Ising nematic (QHIN), has an intrinsic resistive anisotropy for dissipative transport near the center of the corresponding quantum Hall plateau.
\item as a quenched random field is a relevant perturbation to Ising order in $d=2$, the QHIN is unstable to spatial disorder -- such as random potentials or strains -- that gives rise to such fields. Disorder thus destroys the long-range nematic order, giving rise to a paramagnetic phase. Provided that there is (arbitrarily weak) intervalley scattering, this  continues to exhibit the QHE at weak disorder and low temperatures, and is hence termed the quantum Hall random-field paramagnet (QHRFPM). Transport in this phase is dominated by excitations hosted by domain walls between different orientations of the nematic order parameter, and is extremely sensitive to the application of a  symmetry-breaking `valley Zeeman' field  -- for instance, due to uniaxial strain --which can tune between percolating and disconnected domain walls.
\end{enumerate}
Two aspects of this picture are particularly striking and should  apply to a variety of valley quantum Hall ferromagnets. The first is the role of valley anisotropy in establishing the nature of the symmetry breaking. Systems with valleys that are isotropic (for instance, graphene), or have identical anisotropies (such as Si (110) quantum wells) will exhibit an enhanced SU(2)  valley pseudospin symmetry. It is the valley anisotropy in the present situation that entangles rotations in space with those in pseudospin space, and also reduces the order parameter to an Ising variable. Similar behavior is expected for bilayer graphene once trigonal warping of the band structure is included, and for Si (111) heterostructures. Second, we emphasize that the QHIN and the QHRFPM  that naturally emerge in this situation {\it both} exhibit  quantum Hall behavior, but on {\it parametrically different scales}: the QHRFPM shows quantized conductivity only at temperatures below the scale of domain wall-excitations, typically dominated by weak interactions and/or disorder, and hence, much lower than the intrinsic anisotropy scale characteristic of QH transport in the QHIN.

\begin{figure}
  \includegraphics[width=\columnwidth]{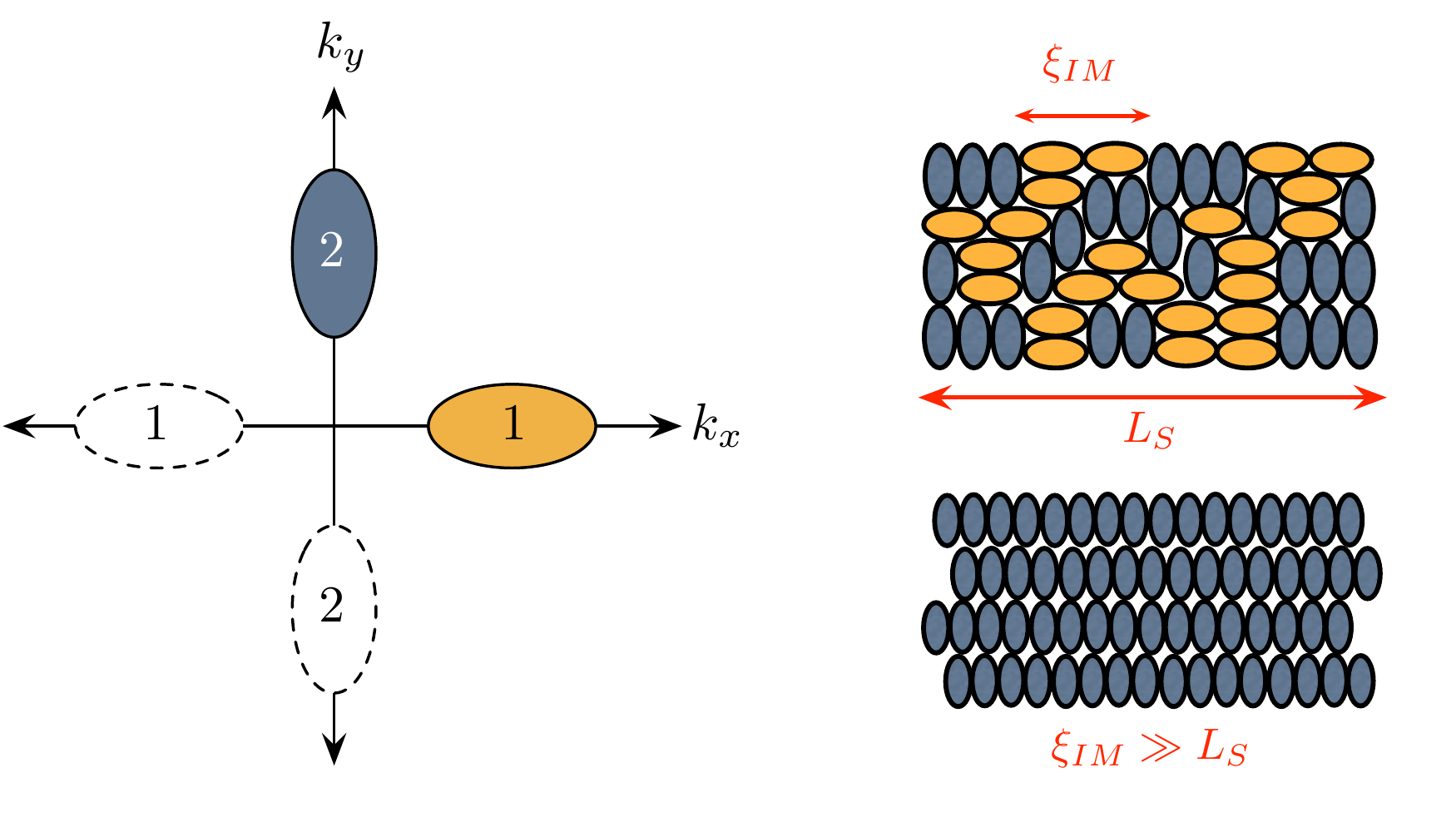}
 \caption{\label{fig-bands}(a.) Model band structure used in this paper, appropriate to describing AlAs wide quantum wells.  (b.) Different phases as determined by comparing  Imry-Ma domain size $\xi_{IM}$ to sample dimensions $L_S$. Top: For $\xi_{IM}\ll L_S$ we find the QHRFPM. Bottom: For $\xi_{IM} \gg L_S$ the system is dominated by the properties of a single domain, and is better modeled as a QHIN. At intermediate scales, $L_S \sim \xi_{IM}$ there is a crossover.}
\end{figure}

A specific example of experimental interest\cite{Shkolnikov:2005p1,Padmanabhan:2010p1,Gokmen:2010p1,Mansour} and our focus in this paper is the case of wide quantum wells in AlAs heterostructures. Here,  two valleys with  ellipsoidal Fermi surfaces are present, as shown in Fig. \ref{fig-bands}. Owing to the anisotropic effective mass tensor in the two valleys, individual electronic states no longer exhibit full rotational invariance. Only discrete rotations of the axes, accompanied by a simultaneous interchange of the valleys remain as symmetries of the system. It is in this specific sense that the internal index is entangled with the spatial symmetries.

The existence of the two phases was originally established within a long-wavelength nonlinear sigma model (\NLSM) field theory, which also provides a caricature of their properties and the above phase diagram in the weak-anisotropy limit. While it is expected that this treatment captures qualitative features of valley Ising physics reasonably well, to make a quantitative connection to experiments a microscopic understanding is essential. Here, we provide such a microscopic analysis of the QHIN, focusing specifically on properties of domain walls which as we have argued are central to this system.

A summary of the main results of this paper, which also serves to outline its organization, follows. We first place this work in context by providing a summary of the important aspects of valley-nematic ordering in the quantum Hall effect in Sec. \ref{sec-overview}, focusing on qualitative features of the phase diagram, the role of thermal fluctuations and quenched disorder and thermal fluctuations, and transport signatures of the QHIN/QHRFPM phases. We then proceed to our technical results. First, we set up a Hartree-Fock formalism (Sec. \ref{sec-HF}), which we  use to obtain a mean-field estimate of the transition temperature out of the thermally disordered phase (Sec. \ref{sec-tc}). We proceed to construct a solution of the HF equations corresponding to a `sharp' domain wall (Sec. \ref{sec-sdw}), where the valley pseudospin changes its orientation abruptly at the wall; this is expected to be an accurate description of physical domain boundaries in the `strongly Ising' limit of large mass anisotropy. We determine the properties of the sharp wall as a function of the mass anisotropy, specifically  its surface tension and dipole moment, the latter a property which is not captured in the \NLSM~limit. We clarify the effect of this dipole moment on critical behavior and domain wall energetics (Sec. \ref{sec-dipoletrans}). We then relax the sharp-wall approximation and numerically solve the HF equations to quantify the amount of `texturing' in a soft domain wall as a function of the anisotropy (Sec. \ref{sec-texture}) -- we note that texturing is a prediction of the \NLSM~that remains valid at high anisotropies. We next turn to an analysis of disorder within the microscopic theory, where we first establish that anisotropies in the screened random impurity potential act as a valley-selection mechanism, translating into a random field acting on the Ising order parameter (Sec. \ref{sec-rfim}), which we compute in Landau-level mixing perturbation theory. We discuss how to estimate the strength of the disorder from the mobility, a measure that is readily accessible to experiments (Sec. \ref{sec-dismob}). Taken together, the domain wall parameters and the random field studies yield estimates for the characteristic domain size due to the disorder, allowing us to make partial contact with experiments (Sec. \ref{sec-exp}). All these results are obtained for the microscopics of the AlAs heterostructures which were the original motivation for our study of valley-nematic order. However, the qualitative features of domain wall structure, random-field disorder, and dipole moment physics apply {\it mutatis mutandis} to other multivalley systems in the QH regime.

\section{\label{sec-overview}Overview: Phases, Transitions, Transport}
\begin{figure}
\includegraphics[width=\columnwidth]{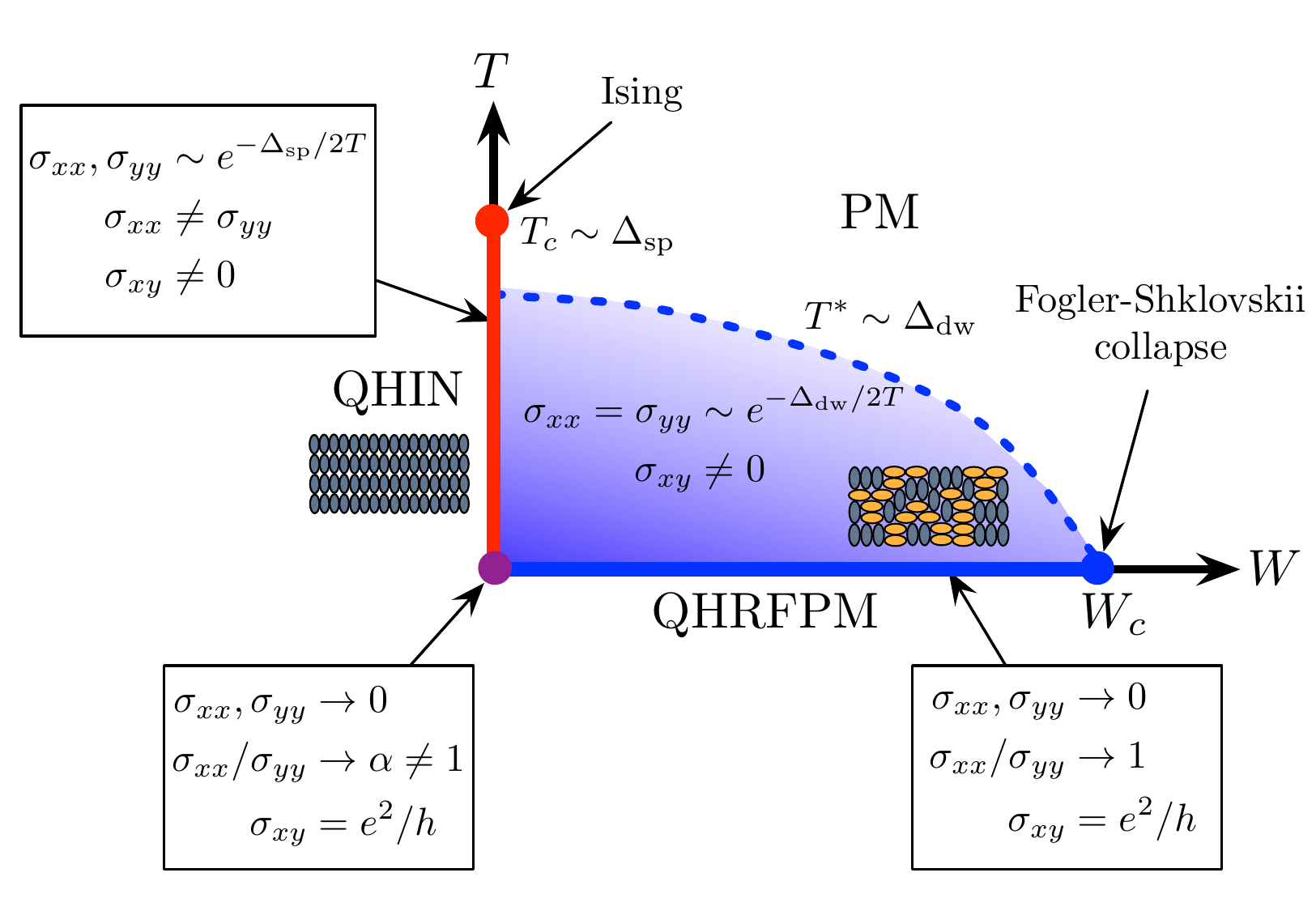}\caption{\label{fig:phasediag} Phase diagram as function of temperature ($T$) and disorder strength ($W$), showing behavior of conductivity. The phases and  critical points are defined in the introduction.}
\end{figure}

The temperature-disorder phase diagram of multivalley 2DEGs exhibiting Ising valley ordering can be sketched as follows (see Fig.~\ref{fig:phasediag}). In the absence of disorder, there is a finite temperature transition into an Ising nematic ordered phase, which exhibits transport features of the QHE. While strictly speaking, the QHE is a zero-temperature phenomenon, in a slight abuse of terminology we will  nevertheless refer to the entire phase below $T_c$ in the zero-disorder limit as the QHIN. The quantization of the Hall conductivity and the vanishing of the longitudinal conductivity are only exponentially accurate at finite temperature. While there is a thermodynamic transition associated with the Ising valley ordering, the conductivity exhibits a  crossover rather than a singularity at $T_c$. The orientational symmetry breaking of the Ising nematic phase is reflected in the anisotropic longitudinal conductivity of the QHIN where $\sigma_{xx}/\sigma_{yy} \neq 1$.
 Upon adding disorder, the Ising transition is destroyed and at $T=0$ the system is in the QHRFPM phase. Above this at finite temperature (shaded region in Fig.~\ref{fig:phasediag}) we once again find exponentially vanishing longitudinal conductivity and exponentially quantized Hall conductivity, but the response is now isotropic: $\sigma_{xx}/\sigma_{yy} = 1$. With similar caveats as in the clean case we will refer to the entire shaded region above the $T=0$ line as the QHRFPM.   In contrast to the QHIN, there is no {\it thermodynamic} phase transition into the QHRFPM at $T>0$, only a crossover in the conductivity at a temperature scale $T^*$ (dashed line in Fig.~\ref{fig:phasediag}.)

 We emphasize  that there is an important qualitative difference between the QHIN and the QHRFPM, over and above the anisotropy in the former. Namely, the crossover into a quantized Hall response in transport is governed by different physical mechanisms. In the QHIN, this crossover occurs at a scale  set by the exchange energy, effectively the single-particle gap, $\Delta_{\text{sp}} \sim e^2/\epsilon\ell_B$ in the QH ferromagnetic ground state. This also sets the scale of the Ising $T_c$, upto a numerical factor that depends on the mass anisotropy. In contrast the QHRFPM is, as we have noted, characterized by multiple domains of differing Ising polarization. Here, the lowest-energy charged excitations are localized on one-dimensional domain boundaries,\cite{*[{For a discussion of the effects of interactions and disorder on transport along domain walls in QHFMs, see }] [{ and }] Falko:1999p1,*Mitra:2003p1} 
 which in the strong-anisotropy limit can be understood in terms of a pair of counterpropagating QH edge states of opposite pseudospin. The stability of the QHE then rests on the gap to creating domain-wall excitations. As this is induced by weak pseudospin symmetry-breaking terms in the Hamiltonian from both disorder and interactions, it is expected to be small and the concomitant conductance quantization is thus fragile. At weak disorder, the dominant source \footnote{While disorder can also lead to scattering between valleys,  this is suppressed owing to the mismatch between the separation of the valleys in momentum space -- roughly an inverse lattice spacing -- and the scale of the random potential fluctuations -- typically several tens of nanometers. Thus, interactions are the dominant source of intervalley scattering in this limit.} of symmetry breaking is from intervalley Coulomb scattering, $V_{iv}$, which thus sets the domain-wall  gap $\Delta_\text{dw}$ and hence the crossover scale $T^*$. For sufficiently strong disorder above a critical strength $W_c$, the energy gap stabilizing the QHRFPM collapses via the  Fogler-Shklovskii scenario\cite{PhysRevB.52.17366} originally devised to describe the collapse of spin-splitting in quantum Hall ferromagnets in GaAs quantum wells.

Whether a particular experimental sample will display the transport anisotropy characteristic of the QHIN, or the isotropic domain-wall dominated transport of the QHRFPM is a matter of quantitative detail, determined by the comparative energetics of the Ising exchange energy and the disorder. Their competition sets a characteristic ``Imry-Ma"\cite{ImryMa} domain size $\xi_{IM}$ in the random-field phase. The question then turns on whether the system consists of a single Ising domain or multiple domains, i.e. it depends on how the domain size compares to the sample dimensions, $L_S$ (see Fig.~\ref{fig-bands}). The exchange strength is determined by the electron-electron interactions, while for the heterostructures of interest the disorder is sensitive to the density of dopant impurities and their typical distance from the plane of the 2DEG. The effective mass anisotropy is important to estimates of both these quantities, for in the isotropic limit there is a full $SU(2)$ pseudospin symmetry, and potential disorder does not exhibit a preference for any particular pseudospin orientation.  Thus, accurate estimates of these quantities  picture are essential to make a quantitative connection with experiments.

\begin{figure}
\includegraphics[width=\columnwidth]{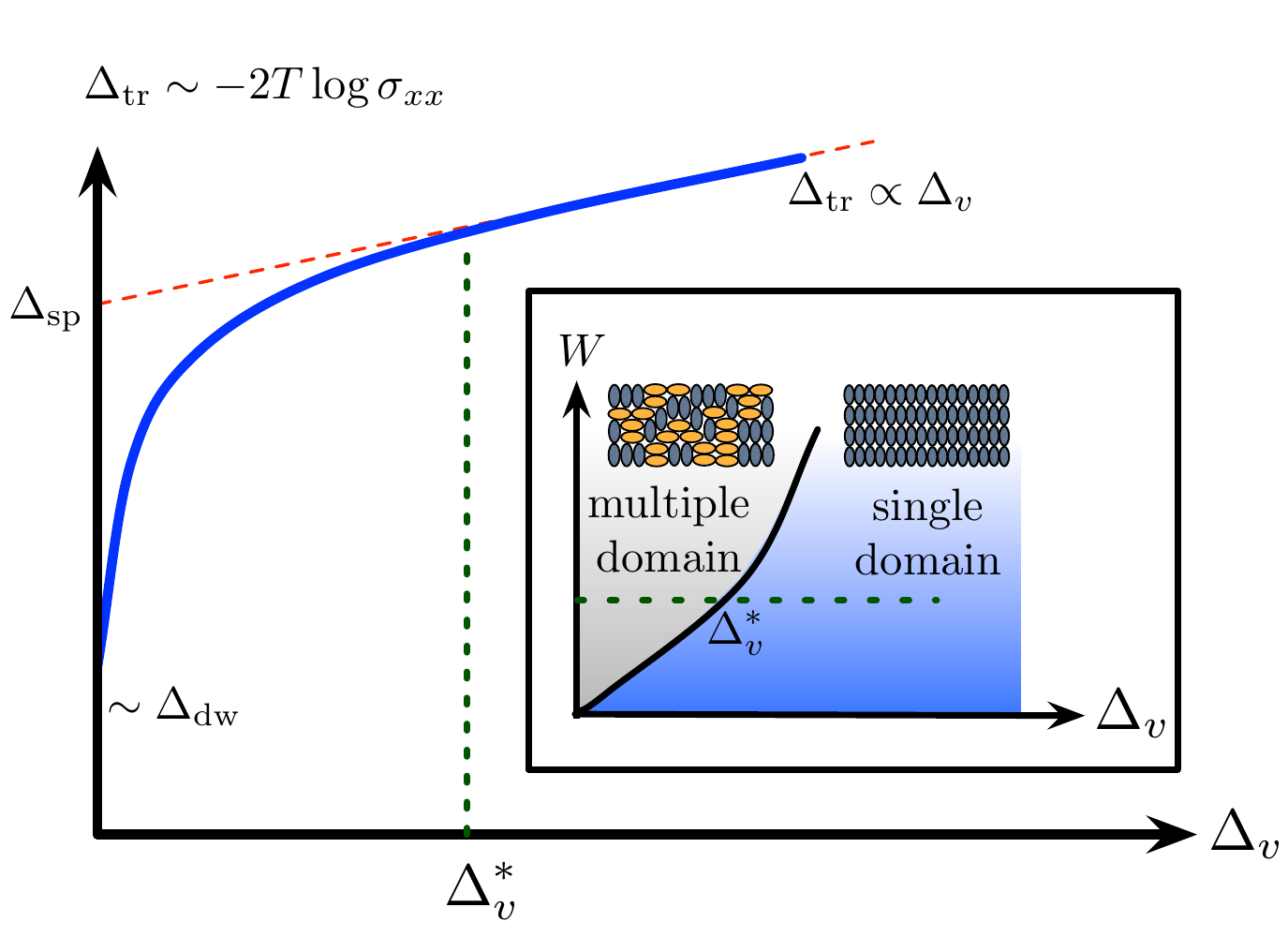}\caption{\label{fig:valleyZeemanresponse} Valley symmetry-breaking field permits transport to probe the energy scales of the QHIN/QHRFPM. (Inset) Domain structure as function of disorder strength and valley splitting; dashed line shows a representative path in $\Delta_v$ leading to a transport signature similar to that in the main figure. $\Delta_v^*$ is the valley splitting for which the system is single-domain dominated.}
\end{figure}

The introduction of an externally applied valley Zeeman field -- experimentally achieved via application of uniaxial strain to the 2DEG -- provides a convenient probe of the transport scales in the QHIN and QHRFPM. First, this field introduces a single-particle splitting between valleys $\Delta_{v}$ that stabilizes the Ising nematic against the effects of disorder. Thus for sufficiently weak disorder and sufficently large $\Delta_v$,  the anisotropic longitudinal conductivity should be clearly established. Second, in the case when for $\Delta_v=0$ the disorder is sufficient that the sample is in the QHRFPM with multiple domains (for instance, along the dotted line in the inset of Fig.~\ref{fig:valleyZeemanresponse}), application of the valley Zeeman field causes a crossover in the the longitudinal conductivity as a function of $\Delta_v$.
 A  sketch of this is provided in Fig.~\ref{fig:valleyZeemanresponse}, and can be understood as follows.
For a disorder strength corresponding to the dotted line in the inset, the system crosses over from multiple domain to single domain behavior. This is reflected in the activation gap for longitudinal transport: in the multiple domain regime, the gap is dominated by the domain wall scale $\Delta_{dw}$. Deep in the single-domain regime, the gap is essentially set by the single-particle gap, which itself scales linearly with $\Delta_v$; the intercept of the asymptotic linear dependence can be used to extract the characteristic single-particle energy scale at zero Zeeman splitting. The sharp crossover between the two regimes can be understood qualitatively in terms of tuning domain walls in a random-field Ising model away from percolation by applying a constant  symmetry-breaking field.

\section{Microscopic Theory}\label{sec-microscopic}

We will begin by developing a microscopic  theory of the QHIN using the Hartree-Fock (HF) approximation.\cite{Fertig:1997p1}
We will focus on the case relevant to AlAs, with two valleys denoted by index $\kappa =1,2$ and centered at $\mathbf{K}_1 = (K_0, 0)$ and $\mathbf{K}_2 = (0,K_0)$ respectively, with mass anisotropy $\lambda^2 = \frac{m_{1,x}}{m_{1,y}} = \frac{m_{2,y}}{m_{2,x}}$ (a schematic dispersion is sketched in Fig.~\ref{fig-bands}.)
 In each valley, the single-particle kinetic energy is
\begin{equation}
T_\kappa = \sum_{i=x,y} \frac{\left(p_i -K_{\kappa,i} + \frac{e}{c}A_i\right)^2}{2 m_{\kappa,i}}
\end{equation}
Working in Landau gauge, $\mathbf{A} = (0,-Bx)$, we find that the lowest LL
 eigenfunctions are
\begin{equation}
\psi_{\kappa, X}(x,y) = \frac{e^{ip_y y}}{\sqrt{L_y\ell_B}} \left(\frac{u_\kappa}{\pi}\right)^{1/4} e^{-\frac{u_\kappa (x-X)^2}{2\ell_B^2}}
\end{equation}

Here, $u_1 = 1/u_2 = \lambda$, and we have labeled states within a LL by their momentum $p_y$, which translates into a guiding-center coordinate via $X = p_y\ell_B^2$.  Henceforth, we shall account for the spatial structure of the LL eigenstates by the standard procedure of projecting the density operators onto the lowest LL. \cite{Moon:1995p1}

\subsection{Hartree-Fock Formalism}\label{sec-HF}

We consider a rectangular system of  dimensions $L_x, L_y$. Since we are interested in a $\nu=1$ state, the total number of electrons in the system (which we take to be even for convenience) is $N = N_{\Phi} =\frac{L_xL_y}{2\pi \ell_B^2} $. For periodic boundary conditions in the $y$-direction, the guiding center coordinate along the $x$-direction is given by $X_n = \frac{2\pi \ell_B^2}{L_y} n$, with $n$ an integer between $-\frac N 2+1$ and $\frac N 2$. For the sake of brevity, we shall continue to label states by $X$, but with the understanding that it is now a discrete index. Unless otherwise mentioned, all sums and products are over the full (finite) range of $X$.

In addition to the electron-electron interaction energy, the lowest Landau level Hamiltonian must include the energy of the electrons interacting with the potential of the positively charged background, which depends on the form of the background charge density. We shall take the positive charges to have orbitals of the same form as electronic states in the two valleys, and corresponding occupation numbers $n^{(b)}_{\kappa}$:
\begin{equation}
\rho_b (\mathbf{r}) =  \sum_{X,\kappa} n^{(b)}_{\kappa} \psi_{X,\kappa}^*(\mathbf{r})\psi_{X,\kappa}(\mathbf{r})
\end{equation}
with $\kappa=1,2$ as before and $n^{(b)}_{1} + n^{(b)}_{2}=1$. Note that for {\it any} choice of $n^{(b)}_{\kappa}$ satisfying the latter constraint, $\rho_b(\br)$ is the {\it same} uniform constant. However, a judicious choice of the background charges will allow us to cancel divergences of the Hartree contribution, as we will see below.

In order to model boundaries between valley domains we also add a spatially varying single-particle pseudospin splitting that increases linearly in $X$ from negative to positive across the system\footnote{Since the number of particles is even, this corresponds to vanishing in between the orbitals at $n=0$ and $n=1$.} which models the external random valley Zeeman field from disorder. This serves a twofold purpose: first, it pins the domain wall\footnote{Without such pinning,  for a finite system the energy optimization would force the domain wall to the boundary where the loss of exchange energy is minimized.} near $n=0$, which is desirable for a stable numerical solution even in the clean limit; second, it allows us to study how the domain wall properties change as we vary the characteristic length scale and typical strength of the random field that leads to domain formation.

With these preliminaries, the second-quantized Hamiltonian projected to the lowest Landau level can now be written in the Landau basis:
\begin{widetext}
\begin{eqnarray}\label{eq:HFham}
H &=& \frac{1}{2} \sum_{\kappa,\kappa'} \sum_{\substack{X,Y \\X',Y'}} V^{\kappa Y, \kappa' Y'}_{\kappa' X', \kappa X} c^\dagger_{\kappa Y}c^\dagger_{\kappa' Y'}c_{\kappa' X'}c_{\kappa X} - \sum_{\kappa}\sum_{X,Y}\left[n^{(b)}_1 V^{\kappa Y, 1X}_{1X, \kappa Y} c^\dagger_{\kappa Y}c_{\kappa Y}+n^{(b)}_2 V^{\kappa Y, 2X}_{2X, \kappa Y} c^\dagger_{\kappa Y}c_{\kappa Y}\right] \nonumber\\& &+{g}   \sum_X\left(\frac{L_y}{2\pi \ell_B^2}X-\frac{1}{2}\right) \left(c^\dagger_{1X} c_{1X} -c^\dagger_{2X} c_{2X} \right) +  E_{\text{self}}\left[\rho_b^2\right]\end{eqnarray}
\end{widetext}
The first term is the electron-electron interaction, the second is the interaction between the electrons and the positive background, and the third term is the single-particle splitting. As discussed in the next section, the characteristic energy scale of this is $\Delta^{\text{SB}}_{d}$, and it varies over a characteristic  distance $d$ corresponding to the correlation length of the random field; rewriting this carefully, leads to the expression given, with ${g} =\frac{\Delta^{\text{SB}}_{d}}{2 d} \frac{2\pi \ell_B^2}{L_y}$ (which has units of energy). Note that because $d\gg\ell_B$, this term is fairly small even at the two ends of the system, where it is maximal. The final term is the self-energy of the background charge distribution, a positive constant that we omit forthwith. In writing  (\ref{eq:HFham}), we have ignored `umklapp' terms that lead to a net transfer of electrons between valleys (as these are exponentially suppressed in $a/\ell_B$, as well as terms that exchange a pair of electrons in the two valleys (suppressed by a factor of $(a/\ell_B)^2$). At the scales of interest, even the latter term only contributes a small energy correction ($\lesssim$ 1\% of the terms kept), and we do not expect their inclusion to significantly alter our conclusions.

The matrix elements of the Coulomb interaction are given by the usual second-quantized form:
\begin{eqnarray}
 V^{\kappa X, \kappa' X'}_{\kappa' Y', \kappa Y} &=& \int d^2r d^2r'\, \psi^*_{\kappa X}(\br)\psi^*_{\kappa' X'}(\br') V(\br-\br') \nonumber\\ & & \times\psi_{\kappa' Y'}(\br')\psi_{\kappa Y}(\br)
\end{eqnarray}
where the single-particle wave functions were defined in the previous section. Note that momentum conservation requires that $X+X' =  Y+Y'$.

\subsubsection{Energy scales}
Throughout the remainder of this paper, we present our results in dimensionless units. We measure energy in units of the Coulomb energy $e^2/\epsilon \ell_B$ where $\epsilon$ is the dielectric constant appropriate to the heterostructure under consideration. For the AlAs devices which are our primary focus, $\epsilon\approx 10$. The magnetic length, $\ell_B \approx 8\,\text{nm}$ for a magnetic field of $10\,\text{T}$, so that $e^2/\epsilon\ell_B \approx 200\,\text{K}$. The surface tension of Ising domain walls is measured in units of $e^2/\epsilon\ell_B^2$, roughly $25\,\text{K}/\text{nm}$ for this choice of parameters.

 \subsection{Estimates of $T_c$}\label{sec-tc}
Our first application of the microscopic theory will be to estimate the Ising ordering temperature $T_c$ for the clean system via finite-temperature Hartree-Fock theory. A standard mean-field decoupling of the Hamiltonian (\ref{eq:HFham}) in the density channel,
$\langle c^\dagger_{\kappa X}c_{\kappa' Y} \rangle = n_{\kappa}\delta_{\kappa\kappa'}\delta_{XY}$, where the occupation numbers are assumed independent of position, yields
\bea
H_{\text{MF}}&=& \frac{1}{2}\sum_{\kappa,\kappa'} \sum_{\substack{X,Y}} \left(n_{\kappa'} -2n^{(b)}_{\kappa'} \right) V^{\kappa Y, \kappa' X}_{\kappa' X, \kappa Y} c^\dagger_{\kappa Y}c_{\kappa Y} \nonumber\\& &\,\,\, - \frac{1}{2}\sum_{\kappa} \sum_{\substack{X,Y}} n_\kappa V^{\kappa Y, \kappa X}_{\kappa Y, \kappa X} c^\dagger_{\kappa Y}c_{\kappa Y}
\eea
We simplify the Hartree term by taking $n^{(b)}_\kappa =  n_\kappa$. For $N\rightarrow \infty$, we may use translation invariance
of the potential to write $H_{\text{MF}} = \sum_{X,\kappa} \epsilon_\kappa c^\dagger_{\kappa X}c_{\kappa X}$,
where
\bea
\epsilon_{\kappa} = -\frac{1}{2}\sum_{\kappa',Y} n_{\kappa'} V^{\kappa X, \kappa' Y}_{\kappa' Y, \kappa X} - \frac{1}{2}\sum_{Y}n_\kappa V^{\kappa X,\kappa Y}_{\kappa Y, \kappa X}
\eea
is independent of $X$.

\begin{figure}
 \includegraphics[width = \columnwidth]{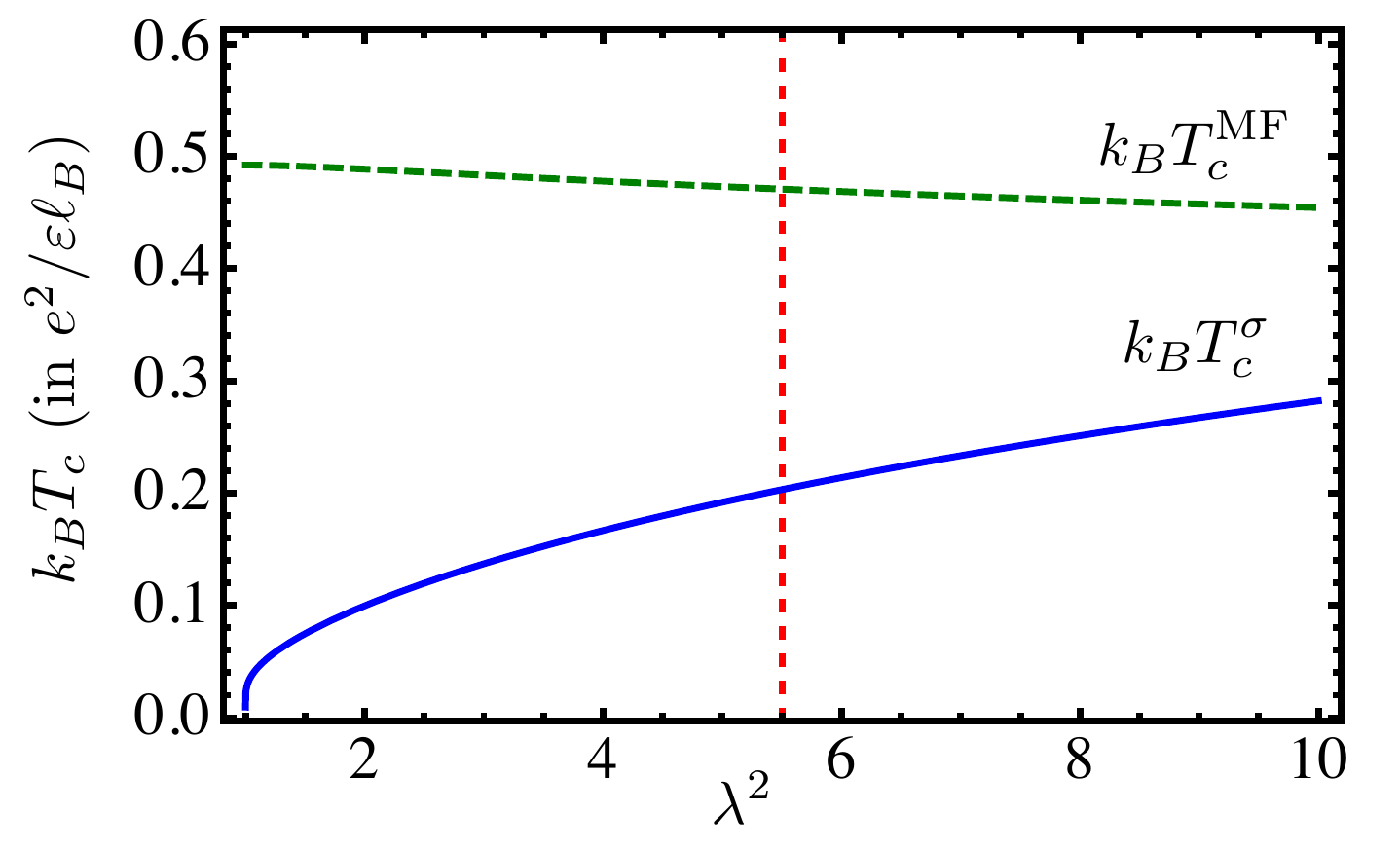}
 \caption{\label{fig-Tc} Mean-field and~\NLSM~estimates of $T_c$. Dashed line shows the anisotropy ($\lambda^2\approx 5.5$) appropriate to AlAs.}
  \end{figure}

We seek a solution where the ground state spontaneously breaks valley symmetry; without loss of generality we may assume it is polarized in valley $1$, and take the energy splitting to be $\Delta$, whence
\bea
n_1 = \frac{e^{\Delta/k_B T}}{1+ e^{\Delta/k_BT}}, \,\,\, n_2 = \frac{1}{1+ e^{\Delta/k_BT}}
\eea
For this ansatz the self-consistency condition corresponds to
$\Delta = \epsilon_2 -\epsilon_1 =  A_1 n_1 - A_2 n_2$, where  after a tedious calculation we find (defining $\bar{1} =2, \bar{2}=1$)
\bea
A_\kappa &=& \frac{1}{2} \sum_{Y}\left[-V^{\bar{\kappa}X, \kappa Y}_{\kappa Y, \bar{\kappa}X}  +V^{\kappa X, \kappa Y}_{\kappa X, \kappa Y} + V^{\kappa X, \kappa Y}_{\kappa Y, \kappa X}\right] \nonumber\\&=& \frac{1}{2}\sum_Y V^{\kappa X, \kappa Y}_{\kappa X, \kappa Y}
\eea
(This cancellation of the Hartree contributions from the two valleys is the reason for the choice of background charge made previously.)
Valley symmetry requires that $A_1=A_2$, which yields the self-consistency condition $\Delta =  A_\kappa \tanh\frac{\Delta}{2 k_BT}$. By the standard comparison of the slope of both sides of this equation at $\Delta=0$, we find for the (mean-field) transition temperature
\bea
k_BT_c^{\text{MF}}= \frac{1}{2} A_\kappa &=&  \frac{1}{4}\sum_Y V^{\kappa X, \kappa Y}_{\kappa X, \kappa Y}  \nonumber\\ &=& \frac{1}{16\pi^2}\frac{e^2}{\epsilon\ell_B}\int_{-\infty}^\infty dx\int_{-\infty}^\infty dy \frac{e^{ -\frac{1}{2} \left( \frac{x^2}{\lambda} + \lambda y^2\right)}}{\sqrt{x^2+y^2}} \nonumber\\ &=& \frac{1}{2(2\pi)^{3/2}} \frac{e^2}{\epsilon\ell_B}\frac{K\left( 1- 1/\lambda^2\right)}{\sqrt{\lambda}}
\eea
where $K$ is the complete elliptic integral of the first kind.\footnote{Note that the $\lambda\rightarrow\lambda^{-1}$ symmetry while not manifest in the final expression for $T_c^\text{MF}$ is nevertheless obtained from an identity satisfied by the elliptic integral $K$.} Note that this mean-field expression for $T_c$ has some unphysical aspects -- most notably it is nonzero even in the Heisenberg limit ($\lambda\rightarrow 1$), and decreases with increasing mass anisotropy. This will be corrected in an RPA spin-wave calculation of quadratic fluctuations about the mean-field ground state. In particular, the fluctuations drive $T_c^\text{MF}$ to zero in the isotropic Heisenberg limit. Furthermore, as spin-wave gap scales roughly with the Ising anisotropy, the debilitating effect of spin waves on $T_c^\text{MF}$ is suppressed at strong anisotropy, offsetting the decrease in the energy scale predicted by the mean-field theory. As the spin-wave calculation is technically involved and not too informative, we provide instead an alternative estimate of $T_c$ for comparison:
 $T_c^{\sigma} \sim 4\pi \rho_s \log^{-1}[\rho_s/\alpha \ell_B^2]$,  obtained from the~\NLSM~with stiffness $\rho_s\approx 0.025\frac{e^2}{16\sqrt{2\pi}\epsilon\ell_B}$ and Ising anisotropy $\alpha \approx 0.01 \frac{e^2}{\epsilon\ell_B^3}(\lambda-1)^2$, whose leading dependence of $\lambda$ was computed in a gradient expansion in Ref.~~\onlinecite{Abanin:2010p1} We plot both estimates in Fig.~\ref{fig-Tc}.

\subsection{Properties of Sharp Domain Walls}\label{sec-sdw}
We turn now to an analysis of `sharp' domain walls. These are solutions to the HF equations where the valley pseudospin abruptly changes orientation from one Landau gauge orbital to the next. We will determine the properties of the sharp domain wall as a function of the anisotropy. While  analytically tractable, this approximation is expected to be a good description of the domain wall only at strong anisotropy, but nevertheless provides a valuable complementary perspective of its properties in a regime where the \NLSM~is no longer valid.
If we take as the ground state a fully pseudospin polarized Slater determinant with all the electrons in valley $1$:
\begin{equation}\label{eq:GSSlater}
\ket{\Psi_G} =\prod_Xc^\dagger_{1X}\ket{0}
\end{equation}
then a  domain wall is captured by a  Slater determinant of the form
\begin{equation}\label{eq:DWSlater}
\ket{\Psi_{\text{DW}}} = \prod_X \left( u_X c^\dagger_{1 X} + v_X c^\dagger_{2 X} \right) \ket{0}.
\end{equation}
The sharp wall corresponds to the case $u_X = 1, v_X = 0$ for $X\leq0$ and $u_X =0, v_x=1$ for $X>$. We once again consider the Hamiltonian (\ref{eq:HFham}) with $g=0$ and assume a background charge distribution polarized in valley 1, i.e. $n^{(b)}_1 = 1$. Two properties of the domain wall will be of especial interest to us: its dipole moment  and its surface tension.

\subsubsection{Surface Tension}\label{sec-sdwst}
\begin{figure}
 \includegraphics[width = \columnwidth]{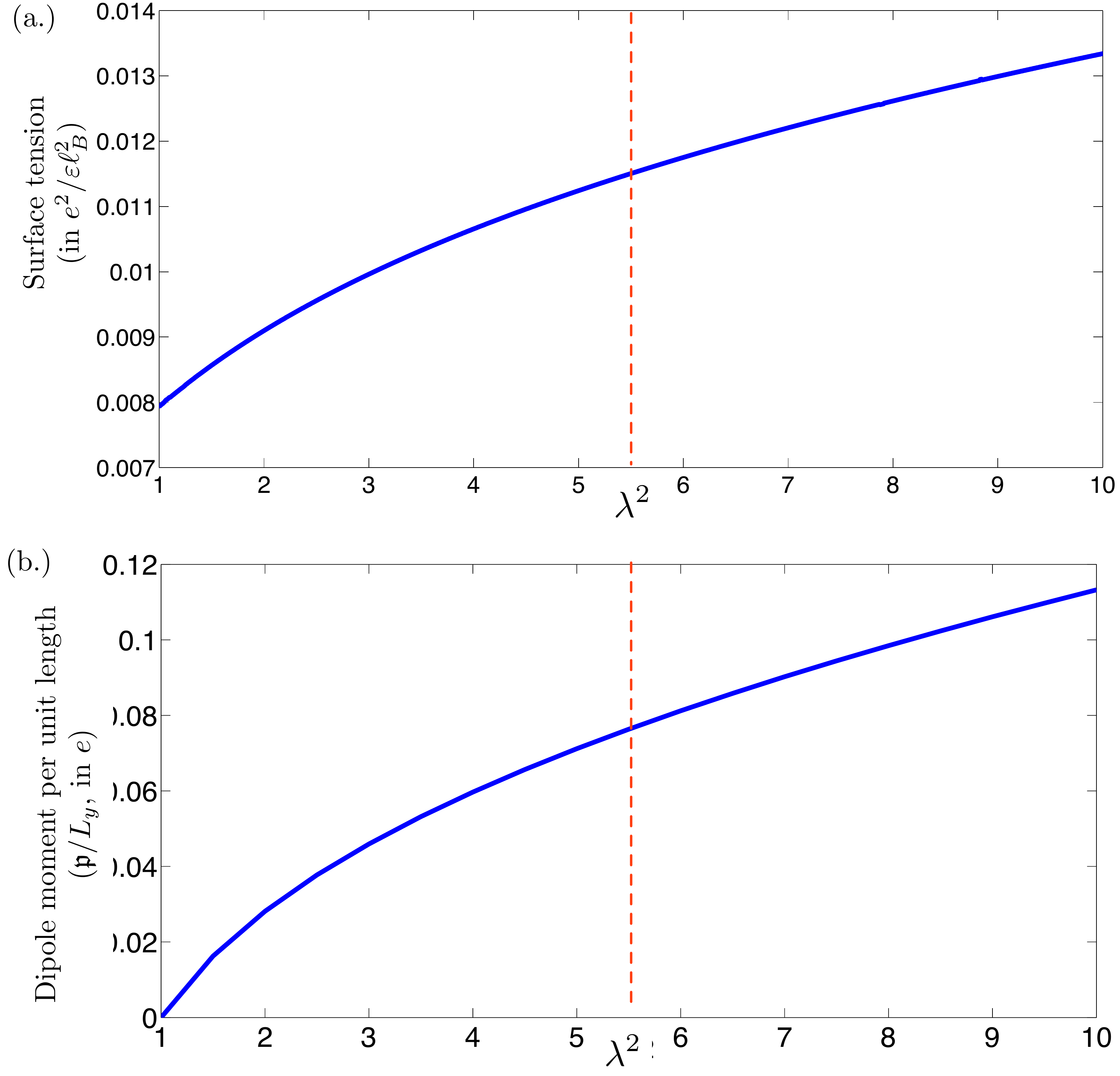}
 \caption{\label{fig-sharpDWresults} (a.) Surface tension and (b.) dipole moment of a sharp DW as a function of the effective mass anisotropy.   Dashed line shows the anisotropy ($\lambda^2\approx 5.5$) appropriate to AlAs.}
  \end{figure}

The first quantity of interest is the domain wall surface tension -- the energy per unit length of the wall. This provides a measure of the Ising exchange energy appropriate to the strong-anisotropy limit. Note that within the \NLSM~the domain wall surface tension depends both on the stiffness and the Ising anisotropy. In the microscopic theory, we find the surface tension (energy per unit  length along the wall) of a sharp domain wall to be the sum of three contributions:
\bea
\sigma(\lambda) &=& \lim_{L_x\rightarrow\infty} \frac{1}{L_y} \left(\langle H \rangle_{\text{DW}} - E_0\right) \nonumber\\&=& E_{\text{I}} + E_{\text{II}} +E_{\text{III}}.
\eea
Here, $E_0$ and $ \left(\langle H \rangle_{\text{DW}} - E_0\right)$  are the energies of the ground state and the sharp domain wall. The three contributions are individually convergent, and can be written as follows. The first term,
\bea
E_\text{I} &=& \frac{1}{2}\sum_{X,X'=1}^\infty \left(V^{1X, 1X'}_{1X',1X}+V^{2X, 2X'}_{2X',2X} - 2V^{1X, 2X'}_{2X', 1X } \right)
\eea
measures the Hartree cost, and can be simplified as
\bea
E_\text{I} &=&\frac{1}{32\pi^2}\frac{e^2}{\epsilon \ell_B^2}\int_{-\infty}^{\infty}dxdx' \int_0^{\frac{L_y}{\ell_B}}dy \frac{f_\lambda(x) f_\lambda(x')}{\sqrt{(x-x')^2+y^2}}\nonumber\\\label{eq:sdwst}
\eea
where 
\bea
f_\lambda(x) = \text{erfc}\left(-x\lambda^{-1/2}\right) -\text{erfc}\left(-x\lambda^{1/2}\right)
\eea
with $\text{erfc}$ the complementary error function. The second term,
\bea
E_{\text{II}} &=& \frac{1}{2}\sum_{X,X'=1}^\infty \left(V^{1X, 1X'}_{1X,1X'}-V^{2X, 2X'}_{2X,2X'} \right)
\eea
is the difference in the `bulk' exchange energy between ground state and domain wall state from orbitals near the center or the edge.The final contribution measures the loss of exchange energy since the two valleys have vanishing exchange matrix elements:
\bea
E_\text{III}=\sum_{X = -\infty}^0\sum_{X' = 1}^\infty V^{1X, 1X'}_{1X,1X'}
\eea
We find $\sigma(\lambda)$ by numerically computing the convergent sums $E_\text{I}, E_\text{II}$ and $E_\text{III}$, in each of which we can truly take the upper bounds on $X$ to infinity.
Note that $\sigma(\lambda)$ depends logarithmically on $L_y$ from the upper bound in the integral in (\ref{eq:sdwst}).   Ignoring this weak dependence, we can take  $L_y\rightarrow\infty$ in integrals over $q_y$ and numerically integrate each term to obtain the surface tension as a function of anisotropy, plotted in Fig.~\ref{fig-sharpDWresults}~(a).

\subsubsection{Dipole Moment}\label{sec-sdwdipole}
Consider for a moment a long-wavelength description of an Ising nematic in terms of a single-component Ising order parameter field $\varphi$, and consider a domain wall parallel to the $y$-axis at $x=0$, between regions with opposite Ising polarization (i.e.,  $\varphi\rightarrow \pm 1$ as $x\rightarrow \mp \infty$.) The remaining rotational symmetry is a rotation $\hat{R}_{\pi}$ that takes $x\rightarrow -x, y\rightarrow -y$. For the given configuration, we have $\hat{R}_\pi \varphi(x,y) = \varphi(-x,-y) = -\varphi(x,y)$. Observe that under this symmetry, $\partial_x \varphi(x,y)$ is left invariant. From this it is not too difficult to show that  $\nabla \varphi(\br)$ transforms as a vector under a rotation by $\pi$, and thus has the same symmetry as that of a dipole moment  normal to the domain boundary and oriented in the direction of decreasing Ising polarization.
It is quite straightforward to find a microscopic origin for the dipole moment. Recall that the spatial extent of the Landau gauge orbitals in the $X$-direction is different in the two valleys. At a domain wall, the charge distribution from valley $1$ decays with a smaller Gaussian envelope than the growth of charge from valley $2$. Assuming a uniform positive background, this leads to a dipole moment associated with the interface between the two valleys and oriented as above. Therefore the theory of an Ising nematic should properly include long-range interactions between dipoles tied to gradients in the Ising order parameter.
However, these appear only at higher orders in the gradient expansion than those used to obtain the leading terms in the long-wavelength theory and represent a small perturbation in the weak-anisotropy limit. It is easier to compute the dipole moment at a domain wall within the microscopic theory: for our choice of background charge, it is straightforward to show that the  charge distribution associated with a sharp domain wall is
\bea
\rho_\text{tot}(\br) &=& \rho_{\text{e}}(\br) +\rho_{\text{bg}}(\br) \nonumber\\
			    &=& - \sum_{X=-N/2+1}^{N/2} \left\{(u_K-1) \psi^*_{1X}(\br)\psi_{1X}(\br)\right.\nonumber\\& & \,\,\,\,\, \,\,\,\,\, \,\,\,\,\,\left.+v_K \psi^*_{2X}(\br)\psi_{2X}(\br) \right\}\nonumber
			    \\&=&  -\sum_{X=1}^{N/2} \left\{\psi^*_{1X}(\br)\psi_{1X}(\br) - \psi^*_{2X}(\br)\psi_{2X}(\br)\right\}.\nonumber\\
\eea
Performing the summations and using the explicit form of the  single-particle wavefunctions, we can verify that $\rho_{\text{tot}}(\br)$ corresponds to a pair of dipolar charge distributions, one located at the domain wall ($X=0$) and the other at the right edge of the system (since the background falls off with a different exponential than the electronic density.) Some care must be taken to separate the contribution of just the dipole at the center so that we have a controlled $L_x \rightarrow\infty$ limit; after some work, we find the dipole moment per unit length of a domain wall is given by
\bea
\frac{1}{L_y} \dipole(\lambda) &=&   \hat{\mathbf{x}}e \left(\lambda - \lambda^{-1}\right)\times \frac{1}{4\pi^{3/2}} \int_{-\infty}^\infty t^2 e^{-t^2} dt    \nonumber\\&=& \frac{e}{8\pi} \left(\lambda - \lambda^{-1}\right)\hat{\mathbf{x}}
\eea
Note that the dipole moment changes sign under $\lambda\rightarrow1/\lambda$, reflecting the fact its existence is directly tied to the   mass anisotropy; we plot this in Fig.~\ref{fig-sharpDWresults}~(b). We reiterate that the dipole moment associated with the DW is a generic feature of an Ising model in which the two phases are distinguished by an orientational symmetry-breaking order parameter; however it is not captured by the \NLSM~ description of Ref.~\onlinecite{Abanin:2010p1} at leading order in the limit of weak anisotropy.

 \subsection{Does the Dipole Moment Matter?}\label{sec-dipoletrans}
A central result of our microscopic study is that there is indeed a nonzero dipole moment at the domain wall as suggested by the symmetries of the system. However, as we have emphasized this physics is invisible in the weak-anisotropy \NLSM~ treatment on the basis of which we sketched the phase diagram of the system with temperature and disorder and discussed qualitative features of these phases. As a consequence of this dipole moment,  there are long-range  interactions between different portions of a domain wall and between different domain walls. Do these perturbations to the original long-wavelength theory affect the physics?
We will address two separate questions: the role they play at the Ising transition in the absence of disorder, as well as the interplay of the long-range couplings with the formation of domains in the Ising phase. As both questions should have  universal answers independent of the microscopic model, it will suffice to consider the role of the dipole-dipole interactions in the long-wavelength theory. Therefore we  consider the free energy of the 2D Ising model,
\bea
\mathcal{F} \sim  \int d^2r \left[ (\nabla \varphi)^2 +   r \varphi^2  + u \varphi^4\right]
\eea
and add to it a perturbation appropriate to a long-range interaction between dipoles:
\bea\label{eq:dipolepert}
\delta\mathcal{F} \sim v\int d^2 r\int d^2 r' \left[\frac{\dipole_\br\cdot \boldsymbol\dipole_{\br'} -(\dipole_\br\cdot \hat{\br})(\dipole_{\br'}\cdot \hat{\br}')   }{|\br -\br'|^3}\right]
\eea
and determine its effect on the critical theory and domain formation with disorder.

\begin{enumerate}[{\it (i)}]
\item {\it Irrelevance at $T_c$.} Using the fact that $\dipole_\br \sim \nabla\varphi(\br)$, we have for dimensional purposes
\bea
\delta\mathcal{F}  \sim v \int d^2 r\int d^2 r' \frac{\varphi(\br) \varphi(\br')}{|\br-\br'|^5}\eea
where we have ignored angular factors as we are really only interested in power-counting. Recall\cite{cardy1996scaling} that a long-ranged spin-spin interaction scaling as $1/x^{d+\sigma}$  is {\it irrelevant} at the short-ranged Ising critical point if $\sigma > 2 - \eta_{\text{SR}}$ where $\eta_{\text{SR}}$ is the anomalous dimension of the Ising field in the short-ranged theory. For dipolar interactions in  the $d=2$ Ising model we have $\eta_\text{SR} = 1/4$ and $\sigma=3$, and thus (\ref{eq:dipolepert}) represents an irrelevant perturbation at the finite-temperature Ising critical point $T_c$.

\item {\it Imry-Ma domain formation at $T=0$.} Recall that  the standard Harris criterion\cite{Harris}/Imry-Ma\cite{ImryMa,PhysRevLett.62.2503} argument in the 2D Ising ordered phase proceeds as follows: we flip spins to orient with the random field to gain an energy  $\propto L$, at the cost of a introducing a smooth domain wall whose energy also scales as $L$; thus, for a sufficiently weak random field there is no advantage to introducing domains. However, a more sophisticated argument\cite{BinderRFIM2D} notes that domain wall roughening can increase the energy gain from the random field so that it scales as $L\log L$. Thus, disorder always destroys the Ising ordered phase in $d=2$. We have verified that long-range dipolar interactions do not affect the qualitative features of this argument, so that disorder remains a relevant perturbation that destroys Ising order at zero temperature.
\end{enumerate}
Although the universal physics and the critical points are unaffected, one physical manifestation of the dipolar interactions is to increase the numerical value of the surface tension and thus renormalize the  Ising stiffness upwards. As a consequence, the characteristic size of an Ising domain in the nematic phase is enhanced  -- note that owing to the exponential dependence of the domain size on the stiffness this can be a quite significant effect.

\subsection{Domain Wall Texturing}\label{sec-texture}
 Thus far we have focused on a sharp domain wall. Within the \NLSM, we find that domain walls are always textured: there is a length scale, set by the competition between the Ising anisotropy (that breaks the $SU(2)$ symmetry down to $\mathbb{Z}_2$) and the stiffness. Does the texturing persist even when the \NLSM~ is no longer valid? We answer this partially via a  self-consistent numerical solution of a domain wall, which reveals that some texturing does indeed persist into the strong anisotropy regime; we also study the texturing as a function of the random field gradient at the wall, as it provides additional information about how the domain wall structure is altered in the presence of disorder.

 We take  (\ref{eq:GSSlater}) as the ground state as before, and the domain wall solution is given by (\ref{eq:DWSlater})
subject now to the constraint $|u_X|^2 + |v_X|^2 =1$, and with the boundary condition that $u_X$ and $v_X$ approach $1$ for $X= (-N/2+1) 2\pi \ell_B^2/L_y$ and $X=N/2\times2\pi \ell_B^2/L_y$ respectively. This corresponds to a domain wall where the pseudospin rotates from valley $1$ to valley $2$ as we move from left to right. Note that, unlike in the sharp case, the wall is allowed to `texture', i.e. cross over from one valley to the other over a finite length scale.  In our simulations, we will take $L_x = 10 \pi \ell_B, L_y= 30 \ell_B$ corresponding to $N=150$, and once again take the background to be fully polarized in valley 1.

Using Wick's theorem and the HF trial wavefunction in (\ref{eq:HFham}), we find
\begin{widetext}
\begin{eqnarray}
\langle H\rangle_{DW} 
=\sum_X\left( \begin{array}{cc} u_X^* & v_X^* \end{array}\right) \left(\begin{array}{cc} U^{\text{H}}_{1}(X) + U^{\text{ex}}_{1}(X) + g\left(\frac{L_yX}{2\pi \ell_B^2} -\frac{1}{2}\right) & U^{\text{c}}(X) \\ U^{\text{c}*}(X) &U^{\text{H}}_{2}(X) + U^{\text{ex}}_2(X) -g\left(\frac{L_yX}{2\pi \ell_B^2} -\frac{1}{2}\right)\end{array}\right) \left( \begin{array}{c} u_X \\ v_X \end{array}\right)\nonumber
\end{eqnarray}
where the Hartree-Fock potentials are
\begin{eqnarray}
U^{\text{H}}_{1}(X) &=&  \sum_{X'}\left[ V^{1X', 1X}_{1X, 1X'} \left(|u_{X'}|^2-1\right)  + V^{2{X'}, 1X}_{1X, 2{X'}} |v_{X'}|^2\right], \,\,\,\,\,
U^{\text{H}}_2(X) =  \sum_Y\left[ V^{1{X'}, 2X}_{2X, 1{X'}} \left(|u_{X'}|^2 -1\right)  + V^{2{X'}, 2X}_{ 2X, 2{X'}} |v_{X'}|^2\right], \nonumber\\
U^{\text{ex}}_1(X) &=&  -\sum_{X'} V^{1{X'}, 1X}_{1{X'}, 1X}|u_{X'}|^2,\,\,\,\,\,
U^{\text{ex}}_{2}(X) =  -\sum_{X'} V^{2{X'}, 2X}_{2{X'}, 2X}|v_{X'}|^2,\,\,\,\,\,
U^{\text{c}}(X) = - \sum_{X'} V^{2{X'} 1X}_{1{X'}, 2X}v_{X'}^* u_{X'}
\end{eqnarray}
\end{widetext}
In the above expressions we have subtracted off the energy of the ground state, so that we may consistently compare domain wall energies for different values of the anisotropy.

The optimization procedure proceeds iteratively, as follows. We begin with a trial wavefunction satisfying the boundary conditions, and in each iteration find the values of $u_m, v_m$ which optimize the HF energy, which are then used to generate the HF potentials for the next iteration. Eventually, the procedure converges to a self-consistent solution.
\begin{figure}
 \includegraphics[width = \columnwidth]{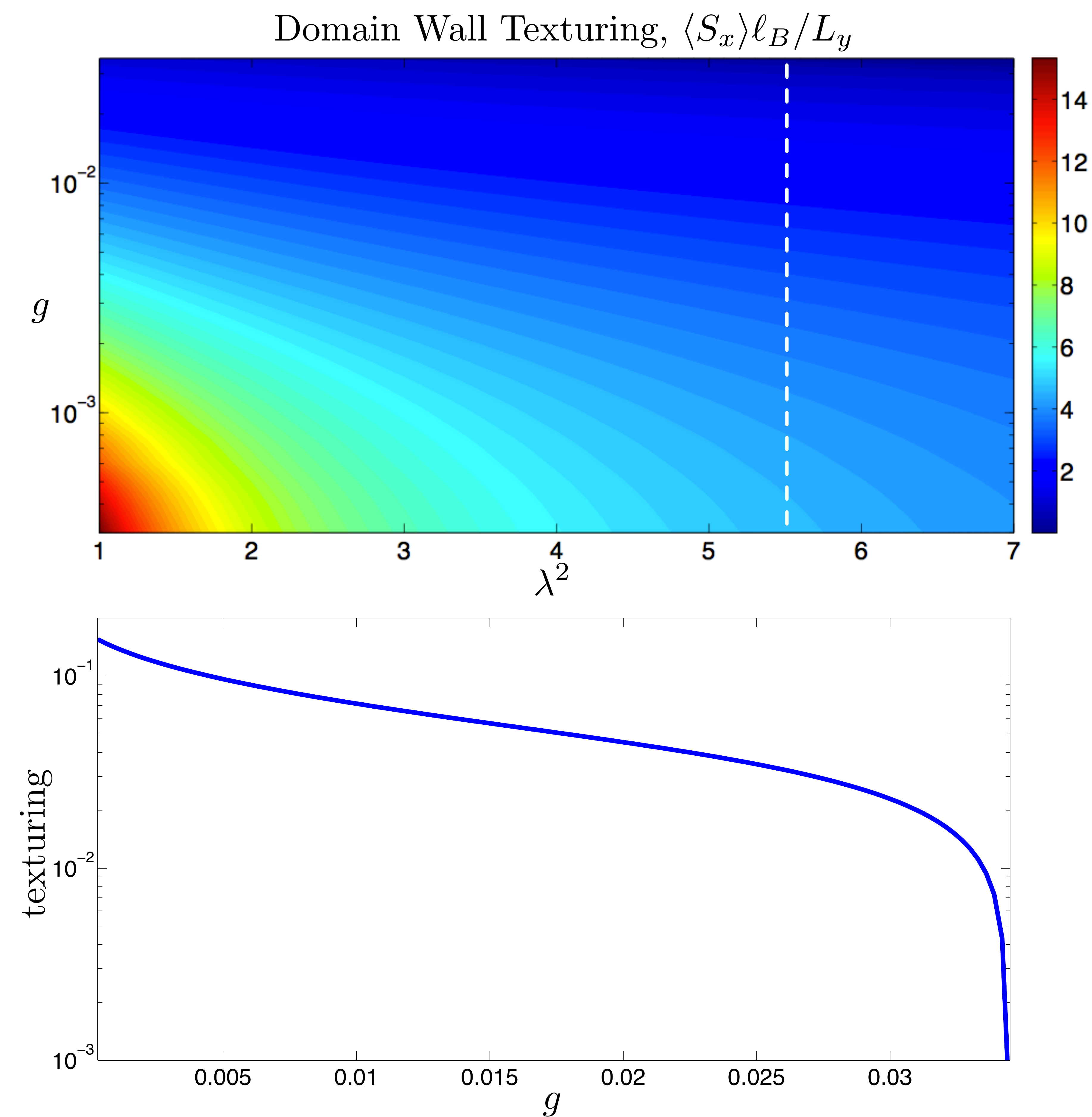}
 \caption{\label{fig-texturing} Domain-wall texturing from Hartee-Fock Theory. (Top) Contour plot of the average in-plane valley pseudospin  $\langle S_x\rangle $ per unit magnetic length along the domain wall, as a function of the mass anisotropy $\lambda^2$ and the valley Zeeman field gradient $g$, with the latter on a logarithmic scale. The dashed line marks the anisotropy $\lambda^2\approx 5.5$ relevant to AlAs; note that there is still some texturing in this limit. (Bottom) Cut along dashed line, with $g$ on a {\it linear} scale.}
\end{figure}

 We estimate the degree of texturing by computing the magnitude of the $x$-component of the pseudospin in the domain wall configuration, since this is nonzero near the wall and vanishes far from it. In Fig. \ref{fig-texturing}, we plot contours of constant $\langle S_{x}\rangle$ in the anisotropy-field gradient plane, as well as the degree of texturing as a function of field gradient at $\lambda^2\approx 5.5$, the anisotropy appropriate to AlAs.

   \section{Disorder in the Microscopic Theory}\label{sec-dis}
As discussed previously, disorder plays a central role in destabilizing the QHIN towards the QHRFPM. There are two primary sources of disorder: (i) random strains in the system can lead to a position-dependent shift of the energies in the two valleys-- while the average strain (pseudomagnetic field) can be externally controlled, fluctuations of the strain are inevitable; and (ii) random fluctuations of the smooth electric potential,
$U_{d}$ that arises from the screening of the potential due randomly placed donor impurities by electrons in the 2DEG also give rise to a random valley field. The random valley Zeeman field from the  strain is difficult to quantify precisely, but is related to the anisotropy of the displacement field $u(\br)$ of the crystal from its equilibrium position: $\Delta_v^\text{str} (\br) \propto (\partial_x -\partial_y) u(\br)$. The random electric field mechanism can be understood via a straightforward application of  perturbation theory and its value estimated from the sample mobility, as we now describe.

\subsection{Random Fields from Impurity Potential Scattering}\label{sec-rfim}
We briefly summarize the argument that leads to a coupling between a local anisotropy in the disorder potential and the Ising order parameter. Since the form factors of the two valleys are different, we expect that the portion of the disorder potential that is antisymmetric in valley indices will lead to a spatially dependent single-particle splitting between valleys; in the limit when the cyclotron gap diverges, i.e. when the lowest Landau level approximation is exact, this is the only contribution, and we can argue from symmetry that the corresponding random field should take the form $(\partial_x^2 -\partial_y^2)U_d$ (at least in the small-anisotropy limit). Note, however, that this term is a total derivative, and contributes significantly only at the boundary of a domain. To go beyond this, we must relax the $\omega_c\rightarrow \infty$ limit, and allow for the effects of Landau-level mixing to first order in $U_d$; since this allows for terms of order $U_d^2/\hbar\omega_c$, the random field now receives contributions of the form $((\partial_xU_d)^2 - (\partial_y U_d)^2)/\hbar\omega_c$, which is not simply a boundary term.

To derive the higher-order contribution to the single-particle valley splitting from the Landau level mixing terms, we make a simplifying assumption: namely, we ignore interactions while computing the effect of mixing. While the interactions may combine with the effects of disorder to modify details of the calculation, we expect that their neglect does not change the qualitative features of our results. The Hamiltonian for noninteracting electrons in AlAs is, in the Landau basis
\begin{eqnarray}
H_{\text{ni}} &=& \sum_{n,{X}} \left(n\hbar\omega_c - \mu\right)c^\dagger_{n,\kappa, X}c_{n,\kappa, X} \nonumber\\
& & + \sum_{\mathbf{q}, {X},\kappa, n,m}U^{mn}_d(-\mathbf{q}) e^{iq_x {X}} c^\dagger_{m,\kappa, X_+}c_{n,\kappa,X_-},
\end{eqnarray}
where we have defined $X_{\pm}= X\pm \frac{q_y\ell_B^2}{2}$. Here,  we have expanded the notation of Section I to include Landau level indices $n$, $m$. In this basis, we have defined the matrix elements of the disorder potential $U_d$ via $U^{mn}_d(-\mathbf{q}) \equiv U_d(-\mathbf{q}) F^{mn}_{\kappa\kappa}(\mathbf{q})
$, which naturally introduces the form factors 
\begin{eqnarray}
 F^{nm}_{\kappa\kappa}(\mathbf{q}) &=& \frac{m!}{n!} \left(\frac{iq_x}{\sqrt{2u_\kappa}} -q_y\sqrt{\frac{u_\kappa}{2}}\right)^{n-m} \nonumber\\& &\times L_{m}^{n-m}\left(\frac{q_x^2}{2u_\kappa} + \frac{q_y^2u_\kappa}{2} \right)e^{-\frac{q_x^2}{4u_\kappa} - \frac{q_y^2u_\kappa}{4}}
\end{eqnarray}
for $n\geq m$, with $F^{nm}_{\kappa\kappa}(\mathbf{q}) = F^{mn}_{\kappa\kappa}(-\mathbf{q})^*$, where $L_{n}^\alpha$ is the generalized Laguerre polynomial.

Next, we compute a renormalized effective potential\cite{HaldaneYang:1997p1} within the lowest Landau level (where $m=n=0$) by including Landau level mixing in perturbation theory. We find
\begin{eqnarray}
\hat{U}_{LLL} &=&  \sum_{\mathbf{q},\bar{X},\kappa} U^{\kappa,{00}}_{d,\text{eff}}(-\mathbf{q})e^{iq_x{X}}c^\dagger_{0,\kappa,X_+} c_{0,\kappa, X_-}
\end{eqnarray}
where, to first order in Landau level mixing,
\begin{eqnarray}
U^{\kappa,00}_{d,\text{eff}}(-\mathbf{q}) &=& U_d(-\mathbf{q}) F^{00}_{\kappa\kappa}(\mathbf{q})  \nonumber\\
& &+ \sum_{\mathbf{q}', n\neq0}\left[ \frac{U_d(-\mathbf{q}')U_d(\mathbf{q}' -\mathbf{q})}{n \hbar\omega_c} e^{i \frac{\mathbf{q}\times\mathbf{q'}\ell_B^2}{2}} \right. \nonumber\\& &\times{F_{\kappa\kappa}^{0n}(\mathbf{q}-\mathbf{q}') F_{\kappa\kappa}^{n0}(\mathbf{q}') } \bigg] +\mathcal{O}\left(|U_d|^3/\hbar\omega_c\right)\nonumber\\
\end{eqnarray}

We are primarily interested in the valley symmetry-breaking contribution from this term, so we consider only the portion antisymmetric in $\kappa$. Assuming that the disorder potential is smooth on the scale of $\ell_B$, we may expand in gradients of $U_d$; to quadratic order in $q_x, q_y$, only the $n=1$ term in the sum contributes, and we find
\begin{eqnarray}
U^{\text{SB}}_{d}(-\mathbf{q}) &=&U^1_{d,\text{eff}}(-\mathbf{q})-U^2_{d,\text{eff}}(-\mathbf{q}) \nonumber\\&=&- \frac{1}{4} \left(\lambda - \lambda^{-1}\right) \left[ \left(\partial_x^2- \partial_y^2\right) U_d\right]_{-\mathbf{q}} \nonumber\\& & + \frac{1}{2\hbar\omega_c}\left(\lambda -\lambda^{-1}\right) \left[\left(\partial_x U_d\right)^2-\left(\partial_y U_d\right)^2 \right] _{-\mathbf{q}}\nonumber\\
\end{eqnarray}
The leading piece vanishes except on domain boundaries, as discussed; thus, the dominant valley splitting arising from impurities is due to the second term.

We focus our attention on a domain boundary, and assume that the distance between the centers of two domains is roughly the correlation length $d$ of $U_d$. In this case, we simply assume that the single-particle energy splitting changes sign linearly over a distance $d$, corresponding to the final term in (\ref{eq:HFham}), with the overall energy scale  $\Delta_{d}^{\text{SB}}$  set by the characteristic scale of the spatially varying random potential $U^{\text{SB}}_d(\br)$.

\subsection{Estimating Disorder Strength from Sample Mobility}\label{sec-dismob}
We may estimate the strength $U_d$ of the smooth random potential from the measured sample mobility $\mu$ and the distance $d$  of the dopant atoms from the plane of the 2DEG, and using the results of the previous section, deduce the parameters of the random Zeeman field $h$. Taking the dopants to be Poisson-distributed, and assuming that the potential fluctuations are screened by electrons in the 2DEG, we can estimate the fluctuations of the potential\cite{Efros:1993p1} in the plane of the 2DEG to be
\begin{equation} \langle |U_d(\mathbf{q})|^2 \rangle =  (U_0d)^2 e^{-2 qd} \end{equation}
where $U_0 $ is determined by the screening length and should be proportional to the impurity density.

The scattering rate due to this potential is
\begin{equation}
 W_{{\bf p}, {\bf p}'}  = {2\pi \over \hbar} | U_d({\bf p} -{\bf p}') |^2 \delta \left(E_{\bf p} - E_{{\bf p}'}\right)
 \end{equation}

A straightforward Boltzmann transport calculation of the transport relaxation time, assuming that it is dominated by the Fermi surface yields
\begin{equation}
{1 \over \tau_{\text{tr}}} = {m \over 2\pi \hbar^2} \int_{-\pi}^{\pi} {d\theta\over 2\pi} \left( 2\pi \over \hbar \right) (1-\cos\theta )U_0^2 d^2 e^{-2k_F \sin {\theta \over 2} \times 2d}
\end{equation}
where  the $(1-\cos\theta)$ factor suppresses the contribution of small-angle scattering, which does not contribute to charge relaxation.
For $k_F d \gg 1$, we have
\begin{equation}{1 \over \tau_{\text{tr}}} =
{m \over \pi \hbar^2}   (U_0 d)^2 {\sqrt{\pi}\hbar^2 \over 8 (m v_F d)^3}.\end{equation}
Using the fact that $1/\tau_{\text{tr}} = e/(m\mu)$ where $\mu$ is the mobility,
\begin{equation}
U_0\approx \left( { 8 \sqrt{\pi} e  \hbar^3 k_F^3 d \over \mu m^2}\right)^{1/2}
\end{equation}
where we take  $m = \sqrt{m_x m_y}$.

Finally, we note that the characteristic length scale of the disorder potential is roughly the distance of the dopant plane from the 2DEG, allowing us to estimate that $|\nabla U_d| \sim U_0/d$. Using the results of the previous section, the characteristic value of the symmetry breaking term $\Delta^{\text{SB}}_d$ is given by
\begin{equation}
\Delta^{\text{SB}}_d\sim {1 \over 2\pi\hbar^2} (m_x -m_y) {\ell_B^4 \over d^2} {U_0^2}.
\end{equation}
This result was used in Ref.~\onlinecite{Abanin:2010p1} where it corresponds to a random field  $h \sim \Delta^{\text{SB}}_{d}/\ell_B^2$ in the \NLSM. Since $h$ is correlated roughly over a distance $d$, we find that the characteristic width of the random field distribution is $W\sim (hd)^2$; this is the parameter that quantifies the strength of disorder in our model.

\section{Experiments}\label{sec-exp}
As promised, we now turn to a discussion of  probes of valley-nematic ordering via transport measurements. We will assume the ability to apply a valley-symmetry-breaking strain. Furthermore, we shall also assume that the maximal valley splitting that can be thus produced is sufficient to fully polarize the system in one of the valleys. We note that this is already feasible for the samples studied experimentally thus far. We will also assume that the sample is engineered in a Hall bar geometry with principal axes parallel to the sample boundaries, so that we may assume that the nematic anisotropy is oriented along the $x$- or $y$- direction of the sample. This removes ambiguity in the definition of components of the conductivity, but more importantly ensures that the anisotropies are observable in the Hall bar geometry.\footnote{This would not be the case, for instance, if one of the principal axes of the Hall bar was oriented along the [110] direction of the quantum well, since the projection of the anisotropic valleys along [100] and [010] onto the [110] direction are identical and thus transport anisotropy no longer reflects valley polarization.}

The cleanest probe of the valley ordering is to examine the  {longitudinal} conductivity for anisotropy. A proxy for the orientational symmetry-breaking order parameter is the quantity $\zeta \equiv \sigma_{xx}/\sigma_{yy} -1$. Note that it is important that both $\sigma_{xx}, \sigma_{yy}$ are measured in simultaneously, which can be conveniently accomplished in a four-terminal geometry..
The behavior of $\zeta$ will exhibit quite distinct behavior as a function of temperature and disorder strength, and will be affected by the application of a strain field. The principal distinction due to disorder is between `clean' samples dominated by the properties of a single Imry-Ma domain, and `dirty' ones which contain several domains. We identify four different cases:
\begin{enumerate}[{\it (i)}]
\item {\it Clean Sample, Zero Strain.} Here, we expect that at high temperatures, the system is in the Ising thermal paramagnet phase, with no anisotropy, so $\zeta=0$; furthermore, $\zeta$ remains flat as the filling is tuned across the Hall plateau. As the temperature is lowered below the Ising $T_c$, the sample should enter the valley-ordered phase. Here, $\zeta$ remains pinned to zero exactly at $\nu=1$ i.e., the center of the Hall plateau. However, upon tuning the filling about $\nu=1$,  $\zeta$ will change sign. This follows from the fact that the longitudinal conductivity goes from being dominated by hopping between hole-like levels of one valley to that between electron-like states of the opposite valley as the doping level crosses the center of the Hall plateau. The resulting longitudinal conductivities inherit the local anisotropy of  Landau orbitals of the two valleys.\cite{Abanin:2010p1,PhysRevB.48.11492} The maximum value attained  for $T\ll T_c$ can be estimated as $\zeta_\text{max} \approx |\sqrt{m_x/m_y}- 1| = \lambda-1$.
\item {\it Clean Sample, Under Strain.} Application of strain to a clean sample should have little effect on the transport below $T_c$ for one orientation of the strain, but should suppress the anisotropy for the opposite orientation. In the paramagnetic phase, a strong valley polarization should result in transport signatures similar to that of the Ising ordered phase.
\item {\it Dirty Sample, Zero Strain.} For dirty samples, the anisotropy from the different domains cancel and we have $\zeta=0$ for zero strain, at all temperatures.
\item {\it Dirty Sample, Under Strain.} Once again, application of strain to a dirty sample should polarize the system, and lead to transport signatures similar to the clean limit at zero strain, below $T_c$. As discussed previously, the activation gap measured via longitudinal transport will be highly sensitive to the application of strain, and increase dramatically as the sample crosses over from multiple-domain to single-domain behavior and thus from domain-wall dominated to single-particle longitudinal transport.
\end{enumerate}
As noted in the preceding section, the Imry-Ma domain size is exponentially sensitive to changes in microscopic parameters and thus estimating the domain size is a challenge. This can be circumvented to some degree by studying transport in samples of different sizes and/or doping levels. For a given doping level, smaller samples are more likely to be in the clean limit as defined above, while lowering the doping level for samples of a fixed size should weaken disorder to some extent. Also, the identification of clean and dirty samples is somewhat loose; samples of intermediate size may show significant anisotropy even though there is no net Ising ordering, since the anisotropies of different domains may not fully cancel.

Note that while the four-terminal probes are particularly unambiguous and striking, there is also useful information that can be gleaned from {\it two-terminal} transport measurements which only have access to a single longitudinal transport coefficient. Here, the nematic symmetry breaking is encoded in the behavior of $\rho_{xx}$ as a function of the doping level. This will be minimal in the center of the Hall plateau, and grow as the filling is detuned from $\nu=1$ in either direction. The mismatch in  $\rho_{xx}$ for $\nu<1$ and $\nu>1$ will exhibit behavior similar to that described for $\zeta$ in the different cases above.

Finally, we note that random field Ising order is typically accompanied by a host of hysteretic effects\cite{Carlson:2006p1}
 that might also be observable in experiments, particularly with an applied valley Zeeman field.

\section{Concluding Remarks}\label{sec-concl}
We have spent the majority of this paper focussing on a specific instance of valley ordering relevant to experiments: the Ising-nematic order in AlAs quantum wells. As the reader no doubt appreciates by now, much of the richness of the phenomena discussed above stems from the inequivalence of the low-energy electronic dispersion in the two valleys.  More specifically, the key observation underpinning our analysis is that the inequivalence between valleys is encoded by the fact that rotating between them necessarily requires a simultaneous interchange of spatial axes; this has three striking consequences. First, in the presence of interactions the na\"ive $SU(2)$ symmetry associated with a generic `internal' index is reduced  to an Ising symmetry. Second, the intertwining of pseudospin and spatial rotations results in the transmutation of quenched {\it spatial} disorder into a random field acting on the Ising order, driving the transition into the paramagnetic QH phase. Finally, the same coupling permits strain to act as a valley Zeeman field, and anisotropy to serve as a probe of transport --- both important to experimental studies of nematic ordering.

This perhaps a good place to observe that other situations in which the valley ordering involves higher symmetry have been studied in the past. A good example is graphene
\cite{*[{See }] [{ for a summary of current experiments and references therein for theoretical studies of QHFM in graphene.}] Young:2012vn}: here, the Dirac dispersion is identical and to good approximation isotropic in the two valleys, and thus the emergent symmetry is $SU(2)$. Another case of historical interest\cite{PhysRevLett.55.433,Rasolt:1986p1} is the (110) surface of Si, the original example of valley QHFM. Although in bulk Si the valleys indeed have substantial anisotropy oriented along different axes, the two valleys that survive in the low-energy dispersion upon projection into the (110) plane have identical anisotropies; therefore, the symmetry here is again $SU(2)$. In these cases, quite different phenomena emerge, such as low-energy skyrmionic `valley textures' and gapless neutral Goldstone modes associated with the breaking of the continuous valley pseudospin symmetry. Furthermore, the equivalence of  the anisotropy in the two valleys strongly diminishes the role of disorder, as it can no longer serve as a valley-selective random field; it thus does not couple directly to the pseudospin index and can act upon it only via the charge sector.

Returning to our central topic, it is clearly desirable to find other instances of Ising-type valley QHFMs. In closing, we would like to flag a few examples as worthy of further study. The first of these, Si (111) heterostructures \cite{KaneValleys2012,PhysRevLett.99.016801}, possess six inequivalent valleys; these split into three pairs, with the dispersion in the two valleys belonging to each pair exhibiting identical anisotropy. Considerations analogous to those presented above suggest that the resulting QHFM should have an $SU(2)\times \mathbb{Z}_3$ symmetry, where the $SU(2)$ rotates between the two valleys within a pair, and the discrete $Z_3$ index acts between the three pairs, once again intertwining spatial and pseudospin rotations. While this is a considerably more intricate symmetry structure than the one considered in this paper, a na\"ive expectation is that now  Ising ordering occurs at $\nu=2$, and involves filling the lowest Landau level in both valleys belonging to a pair thus breaking the $Z_3$ symmetry. While transport measurements suggest that $\nu=2$ does indeed exhibit peculiar behavior,  $\mathbb{Z}_3$ symmetry breaking is more subtle from the point of view of  application of strain and the resulting transport anisotropy. Fully characterizing the symmetry-breaking transition, its experimental consequences, and potential experimental probes, remains an open question. A second example, bilayer graphene, would at first sight appear to exhibit the $SU(2)$ symmetry of its monolayer cousin; however, the inclusion of `trigonal warping' effects into the band structure \cite{PhysRevLett.96.086805} could break this down to an Ising symmetry. Once again, we defer a detailed study of this to future work. Finally, a far more speculative example is the possibility of similar transitions occuring in low-carrier-density systems in three dimensions; recently, transport experiments in Bismuth\cite{Zhu:2012uq,PhysRevB.78.161103, Science} have demonstrated orientational symmetry breaking in the presence of a magnetic field that is not too far from the quantum limit, which could be consistent with some valley-ordering scenarios.\cite{Parameswaran:2012kx}

\acknowledgements{SAP and SLS are grateful to Dmitry Abanin for an earlier collaboration (Ref.~\onlinecite{Abanin:2010p1}), and to Steven Kivelson both for collaboration on that work, as well as insightful suggestions  and comments on the present manuscript.
We acknowledge helpful correspondence with Boris Shklovskii  and
useful discussions with John Cardy, Leonid Glazman, David Huse and Steve Simon, on both that earlier work and the present paper.  We also thank Mansour Shayegan, Tayfun Gokmen, Medini Padmanabhan, Bruce Kane and Tomasz Kott for discussing their experimental data.
We acknowledge support by the Simons Foundation via a Simons Postdoctoral Fellowship at UC Berkeley (SAP), the National Science
Foundation via Grant Number DMR 10-06608 (AK,SLS).}

\bibliography{NematicDW-bib}
\end{document}